\documentclass[conference]{IEEEtran}
\IEEEoverridecommandlockouts
\usepackage{cite}
\usepackage{amsmath,amssymb,amsfonts}
\usepackage{graphicx}
\usepackage{textcomp}
\usepackage{xcolor}
\usepackage{mathtools}
\usepackage{xcolor}
\usepackage{tikz}
\usetikzlibrary{shapes.geometric, arrows.meta, positioning}
\usepackage{algorithm}
\usepackage{algpseudocode}
\usepackage{amsmath}
\usepackage{float}
\def\BibTeX{{\rm B\kern-.05em{\sc i\kern-.025em b}\kern-.08em
    T\kern-.1667em\lower.7ex\hbox{E}\kern-.125emX}}
\begin{document}

\title{MAD-OOD: A Deep Learning Cluster-Driven Framework for an Out-of-Distribution Malware Detection and Classification\\

\thanks{Identify applicable funding agency here. If none, delete this.}
}

\author{\IEEEauthorblockN{1\textsuperscript{st} Tosin Ige}
\IEEEauthorblockA{\textit{Dept. of Computer Science} \\
\textit{The University of Texas at El Paso}\\
Texas, USA \\
toige@miners.utep.edu}
\and
\IEEEauthorblockN{2\textsuperscript{nd} Christopher Kiekintveld}
\IEEEauthorblockA{\textit{Dept. of Computer Science} \\
\textit{The University of Texas at El Paso}\\
Texas, USA \\
cdkiekintveld@miners.utep.edu}
\and
\IEEEauthorblockN{3\textsuperscript{rd} Aritran Piplai}
\IEEEauthorblockA{\textit{Dept. of Computer Science} \\
\textit{The University of Texas at El Paso}\\
Texas, USA \\
apiplai@miners.utep.edu}
\and
\IEEEauthorblockN{4\textsuperscript{th} Asif Rahman}
\IEEEauthorblockA{\textit{Dept. of Computer Science} \\
\textit{The University of Texas at El Paso}\\
Texas, USA \\
arahman3@miners.utep.edu}
\and
\IEEEauthorblockN{5\textsuperscript{th} Olukunle Kolade}
\IEEEauthorblockA{\textit{Dept. of Computer Science} \\
\textit{University of North Carolina}\\
North Carolina, USA \\
ookol@unc.edu}
\and
\IEEEauthorblockN{6\textsuperscript{nd} Sasidhar Kunapuli}
\IEEEauthorblockA{\textit{Independent} \\
\textit{San Jose, CA, USA}\\
sasidhar.kunapuli@gmail.com}

}

\maketitle

\begin{abstract}
Out-of-distribution (OOD) detection remains a critical challenge in malware classification due to the substantial intra-family variability introduced by polymorphic and metamorphic malware variants. Most existing deep learning–based malware detectors rely on closed-world assumptions and fail to adequately model this intra-class variation, resulting in degraded performance when confronted with previously unseen malware families. This paper presents MAD-OOD, a novel two-stage, cluster-driven deep learning framework for robust OOD malware detection and classification. In the first stage, malware family embeddings are modeled using class-conditional spherical decision boundaries derived from Gaussian Discriminant Analysis (GDA), enabling statistically grounded separation of in-distribution and OOD samples without requiring OOD data during training. Z-score–based distance analysis across multiple class centroids is employed to reliably identify anomalous samples in the latent space. In the second stage, a deep neural network integrates cluster-based predictions, refined embeddings, and supervised classifier outputs to enhance final classification accuracy. Extensive evaluations on benchmark malware datasets comprising 25 known families and multiple novel OOD variants demonstrate that MAD-OOD significantly outperforms state-of-the-art OOD detection methods, achieving an AUC of up to 0.911 on unseen malware families. The proposed framework provides a scalable, interpretable, and statistically principled solution for real-world malware detection and anomaly identification in evolving cybersecurity environments.

\end{abstract}


\section{Introduction}
The potency of malware to successfully infiltrate any system no matter how sophisticated made it an indispensable tool available to cybercriminals today, as malware had proven to be highly successful in the extraction of sensitive data which could be used by cybercriminals against their victim. Several machine and deep learning based approaches had been widely proposed  to combat the rampant threat of malware attack but the close-world assumption of identical samples are quickly violated whenever state-of-the-art models are exposed to previously unseen out-of-distribution malware attack. Hence, the persistent proliferation of malicious software (malware) poses an ever-evolving challenge to cybersecurity systems worldwide. With the advent of increasingly sophisticated malware variants and the ubiquity of polymorphic and metamorphic transformations, traditional malware detection mechanisms face significant limitations in achieving high detection accuracy, particularly in the context of out-of-distribution (OOD) detection \cite{zhou2022rethinking,ige2024depth,ige2024towards}. While deep learning and machine learning have emerged as dominant paradigms in malware classification tasks, most state-of-the-art models demonstrate considerable performance degradation when evaluated on previously unseen malware families or variants not represented in their training data \cite{ige2024investigation, ige2024deep, ige2023performance}. This research is motivated by a critical observation: the existing approaches do not effectively exploit the variation in latent embedding spaces between individual variants within the same malware family—a characteristic that is both prevalent and underutilized \cite{ige2022enhancing,ige2022implementation,nguyen2022out,wood2025lmp,park2019generative,ige2023adversarial}.

Malware classification typically relies on the assumption that samples from a particular class provide a comprehensive representation of that class. However, in the domain of malware, this assumption falls short. Each malware family comprises multiple variants, often generated using obfuscation techniques such as encryption, packing, or polymorphism, which drastically alter the feature representations of these samples while preserving their malicious intent \cite{datta2025topology,shafiq2008embedded,kan2021investigating, ige2022ai,ige2024investigation}. Consequently, a single malware variant does not sufficiently represent the diversity of the family in feature space. This insight is a central tenet of our approach, and we argue that it accounts for the poor performance of existing OOD detection methods when applied to malware datasets.


\subsection{Contribution}
The primary contributions of this research are threefold:
\begin{itemize}
  \item\textbf{Novel Framework for OOD Detection:} We propose a two-stage framework for OOD detection and malware classification that exploits the variation in embedding space between variants of the same malware family. The first stage uses spherical decision boundaries defined through Gaussian discriminant analysis to evaluate the likelihood of a test sample being in-distribution, while the second stage integrates this information into a deep neural network for final prediction.

    \item\textbf{Distance-Based Confidence Modeling:} Our method utilizes statistical measures, including Z-score and the coefficient of variation, to determine how far a test sample lies from the centroid of a class distribution. A key insight is the use of multiple z-score comparisons across class boundaries, where a sample must lie within a standard range relative to at least one centroid to be considered in-distribution. This multi-boundary approach accounts for the multidimensional nature of malware distributions.

    \item\textbf{Empirical Validation:} We provide a comprehensive evaluation of our model using malware datasets containing 25 known malware families and multiple out-of-distribution variants. Our model outperforms existing baseline methods in OOD detection, achieving superior AUC scores, particularly for novel malware types.
\end{itemize}
\section{Background and Related Work}
Out-of-distribution detection refers to the task of identifying inputs that do not belong to the distribution of the training data. This is critical for deploying machine learning systems in real-world environments, where they frequently encounter data that differs from what they were trained on. Traditional OOD detection methods in computer vision, such as those based on softmax confidence scores, Mahalanobis distances, and reconstruction error from autoencoders, have demonstrated promising results in standard benchmark datasets like CIFAR-10 and ImageNet \cite{fort2021exploring,yang2022openood,yang2024generalized,adewale2023encoder,okomayin2023ambient}. However, the application of these techniques in the malware domain has been limited and underwhelming due to the distinct characteristics of malware data, including high intra-class variance and distributional overlap among different malware families.

Prior works in malware detection using deep learning have typically employed convolutional neural networks (CNNs) trained on image representations of binaries \cite{wood2025lmp}, recurrent neural networks on opcode sequences \cite{karunanayake2025out}, or graph-based models leveraging control-flow information \cite{um2025spreading}. Although these approaches achieve high classification accuracy under closed-world settings, their performance declines significantly when tasked with recognizing novel, unseen malware families. One-class classification methods, such as One-Class SVM and Support Vector Data Description (SVDD) \cite{ruff2018deep}, have been explored to address this issue \cite{golan2018deep,liang2017enhancing}. SVDD, for example, constructs a hypersphere around training data to capture the normal class distribution and flag samples falling outside the sphere as anomalous. However, these methods often lack scalability or robustness when confronted with the complex embedding distributions exhibited by modern malware datasets.

Recent advances have sought to integrate distance-based anomaly detection techniques into deep learning pipelines. For instance, works like Deep SVDD \cite{lee2018simple} and Geometric Transformations for OOD detection \cite{golan2018deep} attempt to reshape the representation space to better separate in-distribution and out-of-distribution samples. Nonetheless, these methods rarely consider the granularity of variation within individual classes, especially in datasets where class boundaries are inherently fuzzy due to polymorphic behaviors. Our work builds upon this direction by explicitly modeling the intra-class variance within malware families through spherical decision boundaries derived from Gaussian discriminant analysis.

\section{Research Methodology}

\begin{figure}[hbt!]
\centering
\includegraphics[width=1\linewidth]{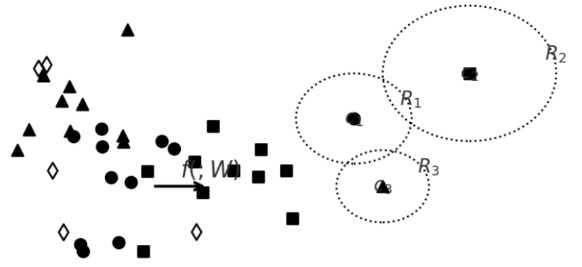}
\caption{\label{fig: Gaussian Discriminant Analysis} Class conditional probability through Gaussian Discriminant Analysis\cite{lee2020multi}}
\end{figure}

\subsection{Spherical Boundary Modeling for Malware Classification}

since each family of malware has different variants leading to variation in the embedding spaces \ref{fig: Malevis and Malimg}. Finding the spherical decision boundary for each class can be instrumental to determines whether a test sample had been previously seen during training or not by integrating the concept of Gaussian discriminant analysis \ref{fig: Gaussian Discriminant Analysis} into deep neural networks. This enables it to learn class-conditional distributions that are explicitly modeled as separable Gaussian distributions using the distance 
of a test sample from each class-conditional distribution to define the confidence score which is eventually used to identifying OOD samples.\\

\begin{equation}
\widehat{L}(u,v) = 
\begin{dcases*}
   1                         & if $u=v$ and $d_v\ne 0$ \\
   -\frac{1}{\sqrt{d_u d_v}} & if $(u,v)\in E$ \\
   0                         & otherwise
\end{dcases*}
\end{equation}\\

To effectively model the various spheres within the dataset, each sphere is treated as distinct class-conditional distributions, specifically assuming isotropic Gaussian distributions to ensure that OOD samples are positioned farther from all predefined class distributions without relying on OOD samples for validation.\\

\begin{equation}
\small
\begin{split}
\mathcal{N}(x \mid \mu_k, \Sigma) =\\
& \frac{1}{(2\pi)^{d/2} |\Sigma|^{1/2}} \exp\left(-\frac{1}{2}(x - \mu_k)^T \Sigma^{-1} (x - \mu_k)\right)
\end{split}
\end{equation}\\

In the proposed approach, an innovative two-stage framework is introduced to address the limitations of current OOD detection methods in malware datasets. The first stage involves modeling class-conditional distributions using spherical decision boundaries derived from Gaussian Discriminant Analysis (GDA) \cite{lee2020multi}. This approach assumes isotropic Gaussian distributions for each class, enabling the estimation of distances between test samples and class centroids to determine likelihood and confidence scores for OOD detection.\\
\\

This approach provides a statistical basis for OOD classification using metrics such as the z-score, which measures the deviation of a sample from the mean in standard deviation units. By using a standard range (e.g., -1 to +1) to classify samples as in-distribution or out-of-distribution, the model effectively introduces a robust, interpretable decision framework. While similar to prior attempts at distance-based classification, this method innovatively combines these statistical insights with deep learning embeddings tailored to the unique characteristics of malware datasets \cite{Ige2024Exploiting,Ige2025Impact}.
This approach is closely related to recent advancements in Deep SVDD [5], which integrates the SVDD loss function into neural network training to learn a compact representation of in-distribution data. Similarly, the proposed method leverages the structure of latent embedding spaces to form well-defined spherical clusters, offering a more granular understanding of family-wise variations.

By measuring the distance between a test sample and the derived class-conditional distributions, we can determine the probability and confidence level of the sample belonging to each class. The theoretical foundation for integrating the distance-based classifier into deep neural networks (DNNs) is provided by Gaussian discriminant analysis (GDA).The qualitative analyses on the latent space obtained by this method provide the strong evidence that it places OOD samples more clearly distinguishable from ID samples in the space compared to the others. The task of data description, also referred to as one-class classification, involves defining a representation of the training data and utilizing it to determine whether a test sample belongs to the same distribution. Earlier research in this area primarily explored kernel-based methods, which sought to establish decision boundaries through identified support vectors. One notable approach, supporting vector data description (SVDD), constructs a closed spherical boundary (hypersphere) that encompasses most of the data.

\subsection{Cluster-Aided Malware Detection and Final Prediction Network}
Following the initial cluster-based classification, the framework employs a second deep learning model that integrates the initial prediction from the clustering algorithm, the output of a supervised DL model, and the input image itself to refine the final prediction. This ensemble-like design enhances the robustness of classification by combining complementary information sources.
Notably, unlike conventional vision-based DL models used in malware detection [6], which typically rely solely on image-to-label mappings, this architecture leverages structural knowledge of the latent space through embedding-based features and prior statistical analysis. This strategy offers a unique advantage in handling polymorphic malware samples, which often defy simple visual patterns.


\begin{figure*}[hbt!]
\centering
\includegraphics[width=1\linewidth]{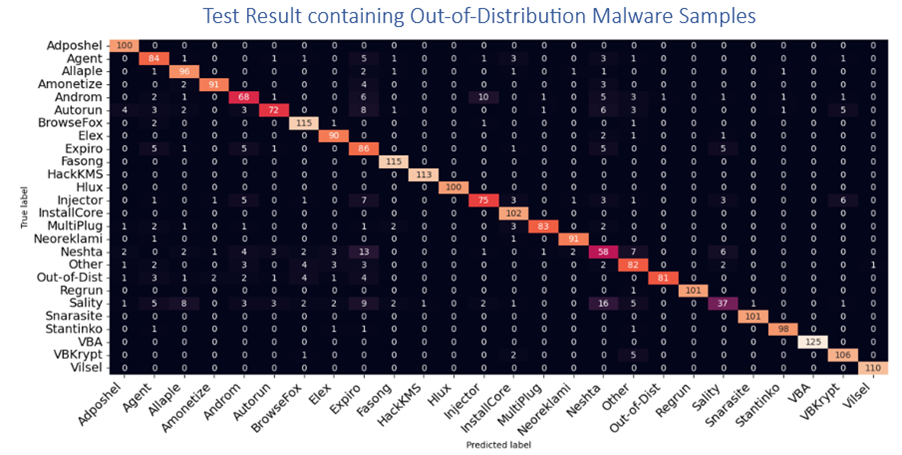}
\caption{\label{fig: Confusion Matrix OOD and IID stage-2} Confusion Matrix on sample test set containing previously unseen out-of-Distribution variants}
\end{figure*}

\begin{table}[hbt!]
\centering
\caption{AUC Score on an Out-of-Distribution Malware Classification} 
\begin{tabular}{|p{0.8in}|p{0.6in}|p{0.8in}|p{0.6in}|}
 \hline
 Malware Family & AUC Score  & Malware Family & AUC Score  \\
 \hline\hline
Adposhel & 0.0059 & InstallCore & 0.539 \\ 
   Agent&0.105&MultiPlug&0.576\\ 
Allaple&0.113&Neoreklami&0.629\\ 
Amonetize&0.166&Neshta&0.585\\ 
Androm&0.266&VBA&0.619\\ 
BrowseFox&0.257&Sality&0.573\\ 
Elex&0.305&Snarasite&0.830\\ 
Expiro&0.360&Stantinko&0.846\\ 
Fasong&0.383& \colorbox{yellow}{Out-of-Dist. (Novel)}&\colorbox{yellow}{0.911}\\ 
HackKMS&0.424&VBKrypt&0.916\\ 
Hlux&0.463&Vilsel&0.993\\ 
Injector&0.499& & \\ \hline\hline
\end{tabular}
\end{table}

\begin{table*}[htbp]
 \caption{Comparison of MAD-OOD with other state-of-the-art Out-of-Distribution models on Benchmark Malware Dataset }
 \begin{center}
 \begin{tabular}{c|c c c c c c c }
 \hline
 \textbf{Method}&\multicolumn{6}{c}{\textbf{Model Comparison and Evaluation}} \\
 \cline{2-7} 
 \textbf{ } & \textbf{\textit{AUC}}& \textbf{\textit{AP-Id}}& \textbf{\textit{AP-OOD}}& \textbf{\textit{FPR}}& \textbf{\textit{AR-OOD}} &\textbf{\textit{ACC}}  \\
 \hline
 MSP& 0.611& 0.464& 0.322& 0.613& 0.526& 53.82 \\
 \hline
 OE& 0.247& 0.634&0.709 & 0.751 &0.592 & 0.407\\
 \hline
 EnergyOE& 0.651& 0.660& 0.792& 0.736&0.808 &0.682 \\
 \hline
 OCL& 0.637& 0.529&0.558 &0.771 &0.690 & 0.625 \\
 \hline 
 PASCL & 0.209& 0.405 &0.392& 0.229& 0.393 &0.592 \\
 \hline
 OS & 0.692& 0.442 &0.842& 0.793&0.712 &0.827 \\
 \hline
 Class Prio & 0.596& 0.376 &0.816& 0.728&0.693 &0.606 \\
 \hline
 BERL & 0.846& 0.572 &0.561& 0.807&0.737 &0.812 \\
 \hline
 MAD-OOD & 0.906& 0.951 &0.861& 0.940&0.817 &0.907 \\
 \hline
 \multicolumn{4}{l}{$^{\mathrm{a}}$}
 \end{tabular}
 \label{tab1e}
 \end{center}
 \end{table*}

\begin{table*}[htbp]
\caption{Model Evaluation on Benchmark Malware Dataset}
\begin{center}
\begin{tabular}{c|c c c c c c c }
\hline
\textbf{OOD Malware}&\multicolumn{6}{c}{\textbf{Proposed Model Generalization}} \\
\cline{2-7} 
\textbf{Dataset} & \textbf{\textit{AUC}}& \textbf{\textit{AP-Id}}& \textbf{\textit{AP-OOD}}& \textbf{\textit{TPR}}& \textbf{\textit{AR-OOD}} &\textbf{\textit{ACC}}  \\
\hline
MaleVis& 0.911& 0.864& 0.822& 0.813& 0.926& 93.82 \\
\hline
BODMAS& 0.847& 0.834&0.809 & 0.851 &0.792 & 0.907\\
\hline
Virus-MNIST& 0.851& 0.860& 0.792& 0.936&0.908 &0.882 \\
\hline
Stamina& 0.837& 0.929&0.858 &0.871 &0.890 & 0.925 \\
\hline MalImg & 0.906& 0.951 &0.861& 0.940&0.817 &0.907 \\
\hline
\multicolumn{4}{l}{$^{\mathrm{a}}$Summary of OOD-MAD model against benchmark malware dataset}
\end{tabular}
\label{tab1}
\end{center}
\end{table*}

\section{Metric Evaluation}
Our metric evaluation is based on the centroid of each decision boundary as I can now measure the distance between a test sample and the derived class-conditional distributions and utilize it to determine the probability and confidence level of any particular sample belonging to a class to determine whether it is an in-distribution or out-of-distribution sample. In this, utilizing the class conditional distributions to determine the centroid of each decision boundaries. After this comes the challenge of how far new data points could be to any of the centroid before classifying it as in-distribution or an out-of-distribution sample.

\begin{figure}[hbt!]
\centering
\includegraphics[width=1\linewidth]{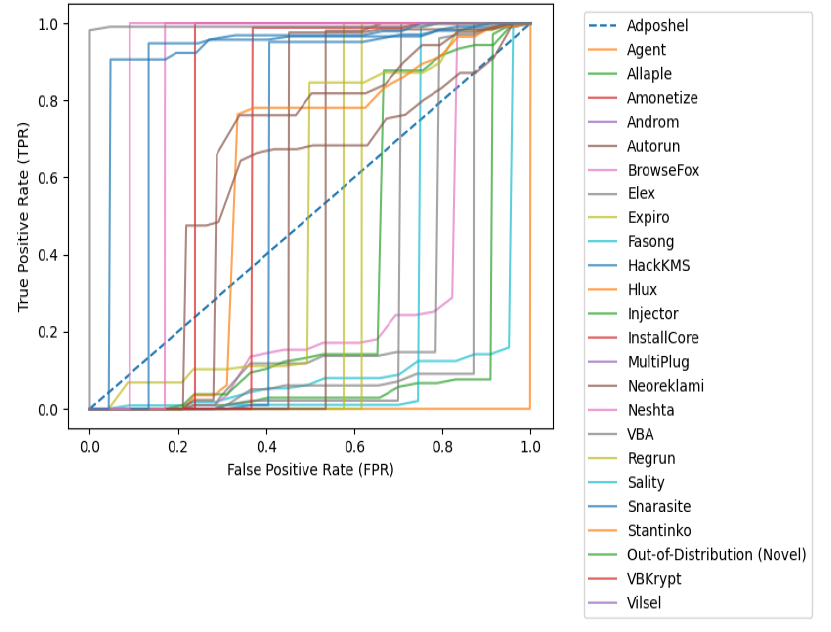}
\caption{\label{fig: Confusion Matrix OOD and IID} ROC Performance Evaluation of the Proposed Framework on the prediction of Previously Unseen Out-of-Distribution Classification}
\end{figure}

We computed the standard deviation of data point from any new samples to see how far apart it is to each of the centroid of the spherical boundary, the results was that of a low standard deviation between data points of news sample to the centroid and high standard deviation for others, then arises the problem of no single rule to determine a low standard deviation across several centroids. This lead to computation of the CV (coefficient of variation) which is the ratio of the distance between the mean and standard deviation considering it has specific range to determine and outliner. Z-score as a statistical method can detect outliers by measuring how far a data point deviates from the mean in terms of standard deviations using the formula:

\begin{equation}
z = \frac{x - \mu}{\sigma}
\end{equation}

\noindent
\textbf{Where:}
\begin{itemize}
  \item $z$ z-score (the number of standard deviations a data point is from the mean)
  \item $x$ is the observed value
  \item $\mu$ is the mean of the data point within the cluster
  \item $\sigma$ is the standard deviation of the new data point to the centroid cluster standard deviation
\end{itemize}

\begin{itemize}
  \item\textbf{Computation of the Mean and Standard Deviation:}Calculate the mean ($\mu$) and standard deviation ($\sigma$) of the selected feature across a dataset of benign and potentially malicious samples.

\item\textbf{Z-score computation:}Determine the Z-score for each family represented by the centroid.

\item\textbf{Pre-Determine in-distribution Threshold:} Common thresholds for outlier detection

Z > 1 (or Z < -1) → Possible outlier

Z > 2 → Highly suspicious

\item\textbf{Initial Input Classification: }If a sample has a Z-score beyond the chosen threshold, it is flagged as an outlier and so is initially classify a possible out-of-distribution sample

\end{itemize}

So, to compare new data point from the test set, I use z-score because it has a standard range to determine an outlier data point.
z-score (z) = (new data point - mean) / standard deviation
The standard to determine outlier in z-score is that anything outside the range of -1 and +1 is an outliner. Hence, for any sample to be classify as an in-distribution sample, at least one of the z-score to each of the centroid of the spherical boundary must be within the range of -1 and +1.
If all of the z-scores falls outside the -1 and +1 range, then it is consider as an out of distribution sample.
If the z-score is less than -1 or greater than 1, the data point is an outlier as it is further away from the mean than majority of the other data points.

\subsection{Training a Computer Vision Based-Deep Learning Classifier for Initial family Prediction}

\textbf{Splitting Data}
Train, validation, and test sets are carefully divided into:
Train \& Validation Sets: Contain only in-distribution (known malware and benign) data along with sample representation from out-of-distribution set.
Test Set: Contains both in-distribution and OOD samples for evaluation.

\textbf{Training Procedure}
Step 1: Train on In-Distribution Data
Loss Function: Use cross-entropy for classification (malware vs. benign).
Optimizer: Adam and SGD with learning rate scheduling.
Regularization: we applied dropout, and batch normalization

\textbf{Evaluation Metrics}
In-Distribution Performance: Accuracy, and Confusion matrix for malware classification.
OOD Detection Performance: AUROC (Area Under ROC Curve): Measures separation between in-distribution and OOD

\subsection{Training a Computer Vision Based-Deep Learning Classifier for Final Prediction}

Second Model is a deep neural network-based model which takes three arguments
Argument 1-prediction from the cluster analysis
Argument 2: prediction output from the first model
Argument 3: Input image
Unlike the first model, we make use of the new embedding spaces from the spherical boundary to train the second model since the final prediction comes from the second model. There are two stages to it here
i. Initial prediction from the second model, since the second model is train on the new embedding spaces, We can expect a better predictive output from it
ii.  z-score Calculation

While previous work has explored embedding-based OOD detection [3][4], few have addressed the unique challenges posed by malware datasets. This study fills a critical gap by acknowledging and explicitly modeling the intra-family variation in malware, which is often overlooked in favor of simplifying assumptions. The integration of GDA-inspired decision boundaries, z-score analysis, and a multi-input deep learning classifier offers a holistic approach that outperforms existing methods in separating in-distribution malware from novel, unseen threats.
Moreover, the empirical results support the efficacy of the proposed method. For instance, the Area Under the Curve (AUC) scores for known malware families such as Regrun (0.761), Snarasite (0.830), and Stantinko (0.846) indicate improved separability in the latent space, while the detection score for novel OOD samples reaches up to 0.911—highlighting its practical potential in real-world malware detection.

\section{Contributions Over Prior Work}
This research introduces a novel two-stage framework for out-of-distribution (OOD) detection in malware classification by exploiting the inherent variability among malware variants within the same family. Unlike traditional models that assume class uniformity, this approach integrates unsupervised clustering with deep neural networks and Gaussian discriminant analysis (GDA) to establish spherical decision boundaries around malware family embeddings, effectively identifying unseen samples without requiring OOD data during training. A z-score–based statistical analysis enhances discrimination between in-distribution and outlier data. The second stage refines classification using enriched embeddings and multi-source inputs, resulting in superior predictive and OOD detection performance. Achieving an AUC score of 0.911, the model outperforms conventional methods, offering a robust, scalable, and statistically grounded solution for accurate malware detection, with potential extensions to real-time cybersecurity applications.

%


\bibliographystyle{plain}
\bibliography{references.bib}

@article{fort2021exploring,
  title={Exploring the limits of out-of-distribution detection},
  author={Fort, Stanislav and Ren, Jie and Lakshminarayanan, Balaji},
  journal={Advances in neural information processing systems},
  volume={34},
  pages={7068--7081},
  year={2021}
}

@article{yang2022openood,
  title={Openood: Benchmarking generalized out-of-distribution detection},
  author={Yang, Jingkang and Wang, Pengyun and Zou, Dejian and Zhou, Zitang and Ding, Kunyuan and Peng, Wenxuan and Wang, Haoqi and Chen, Guangyao and Li, Bo and Sun, Yiyou and others},
  journal={Advances in Neural Information Processing Systems},
  volume={35},
  pages={32598--32611},
  year={2022}
}

@article{yang2024generalized,
  title={Generalized out-of-distribution detection: A survey},
  author={Yang, Jingkang and Zhou, Kaiyang and Li, Yixuan and Liu, Ziwei},
  journal={International Journal of Computer Vision},
  volume={132},
  number={12},
  pages={5635--5662},
  year={2024},
  publisher={Springer}
}

@inproceedings{zhou2022rethinking,
  title={Rethinking reconstruction autoencoder-based out-of-distribution detection},
  author={Zhou, Yibo},
  booktitle={Proceedings of the IEEE/CVF Conference on Computer Vision and Pattern Recognition},
  pages={7379--7387},
  year={2022}
}

@inproceedings{ige2024depth,
  title={An in-Depth Investigation Into the Performance of State-of-the-Art Zero-Shot, Single-Shot, and Few-Shot Learning Approaches on an Out-of-Distribution Zero-Day Malware Attack Detection},
  author={Ige, Tosin and Kiekintveld, Christopher and Piplai, Aritran and Wagler, Amy and Kolade, Olukunle and Matti, Bolanle Hafiz},
  booktitle={2024 International Symposium on Networks, Computers and Communications (ISNCC)},
  pages={1--6},
  year={2024},
  organization={IEEE}
}

@inproceedings{ige2024towards,
  title={Towards an in-depth evaluation of the performance, suitability and plausibility of few-shot meta transfer learning on an unknown out-of-distribution cyber-attack detection},
  author={Ige, Tosin and Kiekintveld, Christopher and Piplai, Aritran and Wagler, Amy and Kolade, Olukunle and Matti, Bolanle Hafiz},
  booktitle={2024 International Symposium on Networks, Computers and Communications (ISNCC)},
  pages={1--6},
  year={2024},
  organization={IEEE}
}

@inproceedings{ige2024investigation,
  title={An investigation into the performances of the state-of-the-art machine learning approaches for various cyber-attack detection: A survey},
  author={Ige, Tosin and Kiekintveld, Christophet and Piplai, Aritran},
  booktitle={2024 IEEE International Conference on Electro Information Technology (eIT)},
  pages={135--144},
  year={2024},
  organization={IEEE}
}

@article{golan2018deep,
  title={Deep anomaly detection using geometric transformations},
  author={Golan, Izhak and El-Yaniv, Ran},
  journal={Advances in neural information processing systems},
  volume={31},
  year={2018}
}

@inproceedings{ruff2018deep,
  title={Deep one-class classification},
  author={Ruff, Lukas and Vandermeulen, Robert and Goernitz, Nico and Deecke, Lucas and Siddiqui, Shoaib Ahmed and Binder, Alexander and M{\"u}ller, Emmanuel and Kloft, Marius},
  booktitle={International conference on machine learning},
  pages={4393--4402},
  year={2018},
  organization={PMLR}
}

@article{liang2017enhancing,
  title={Enhancing the reliability of out-of-distribution image detection in neural networks},
  author={Liang, Shiyu and Li, Yixuan and Srikant, Rayadurgam},
  journal={arXiv preprint arXiv:1706.02690},
  year={2017}
}

@article{lee2018simple,
  title={A simple unified framework for detecting out-of-distribution samples and adversarial attacks},
  author={Lee, Kimin and Lee, Kibok and Lee, Honglak and Shin, Jinwoo},
  journal={Advances in neural information processing systems},
  volume={31},
  year={2018}
}

@article{wood2025lmp,
  title={LMP-GAN: Out-of-Distribution Detection for Non-Control Data Malware Attacks},
  author={Wood, David and Kapp, David and Kebede, Temesgen and Hirakawa, Keigo},
  journal={IEEE Transactions on Pattern Analysis and Machine Intelligence},
  year={2025},
  publisher={IEEE}
}

@article{karunanayake2025out,
  title={Out-of-distribution data: an acquaintance of adversarial examples-a survey},
  author={Karunanayake, Naveen and Gunawardena, Ravin and Seneviratne, Suranga and Chawla, Sanjay},
  journal={ACM Computing Surveys},
  volume={57},
  number={8},
  pages={1--40},
  year={2025},
  publisher={ACM New York, NY}
}

@inproceedings{um2025spreading,
  title={Spreading Out-of-Distribution Detection on Graphs},
  author={Um, Daeho and Lim, Jongin and Kim, Sunoh and Yeo, Yuneil and Jung, Yoonho},
  booktitle={The Thirteenth International Conference on Learning Representations},
  year={2025}
}

@article{datta2025topology,
  title={Topology of Out-of-Distribution Examples in Deep Neural Networks},
  author={Datta, Esha and Hennig, Johanna and Domschot, Eva and Mattes, Connor and Smith, Michael R},
  journal={arXiv preprint arXiv:2501.12522},
  year={2025}
}

@inproceedings{nguyen2022out,
  title={Out of distribution data detection using dropout bayesian neural networks},
  author={Nguyen, Andre T and Lu, Fred and Munoz, Gary Lopez and Raff, Edward and Nicholas, Charles and Holt, James},
  booktitle={Proceedings of the AAAI Conference on Artificial Intelligence},
  volume={36},
  number={7},
  pages={7877--7885},
  year={2022}
}

@inproceedings{park2019generative,
  title={Generative malware outbreak detection},
  author={Park, Sean and Gondal, Iqbal and Kamruzzaman, Joarder and Oliver, Jon},
  booktitle={2019 IEEE International Conference on Industrial Technology (ICIT)},
  pages={1149--1154},
  year={2019},
  organization={IEEE}
}

@inproceedings{shafiq2008embedded,
  title={Embedded malware detection using markov n-grams},
  author={Shafiq, M Zubair and Khayam, Syed Ali and Farooq, Muddassar},
  booktitle={International conference on detection of intrusions and malware, and vulnerability assessment},
  pages={88--107},
  year={2008},
  organization={Springer}
}

@inproceedings{kan2021investigating,
  title={Investigating labelless drift adaptation for malware detection},
  author={Kan, Zeliang and Pendlebury, Feargus and Pierazzi, Fabio and Cavallaro, Lorenzo},
  booktitle={Proceedings of the 14th ACM Workshop on Artificial Intelligence and Security},
  pages={123--134},
  year={2021}
}

@inproceedings{lee2020multi,
  title={Multi-class data description for out-of-distribution detection},
  author={Lee, Dongha and Yu, Sehun and Yu, Hwanjo},
  booktitle={Proceedings of the 26th ACM SIGKDD International Conference on Knowledge Discovery \& Data Mining},
  pages={1362--1370},
  year={2020}
}

@inproceedings{ige2023performance,
  title={Performance comparison and implementation of bayesian variants for network intrusion detection},
  author={Ige, Tosin and Kiekintveld, Christopher},
  booktitle={2023 IEEE International Conference on Artificial Intelligence, Blockchain, and Internet of Things (AIBThings)},
  pages={1--5},
  year={2023},
  organization={IEEE}
}

@article{ige2024deep,
  title={Deep learning-based speech and vision synthesis to improve phishing attack detection through a multi-layer adaptive framework},
  author={Ige, Tosin and Kiekintveld, Christopher and Piplai, Aritran},
  journal={arXiv preprint arXiv:2402.17249},
  year={2024}
}

@article{ige2022ai,
  title={AI powered anti-cyber bullying system using machine learning algorithm of multinomial na{\"\i}ve Bayes and optimized linear support vector machine},
  author={Ige, Tosin and Adewale, Sikiru},
  journal={arXiv preprint arXiv:2207.11897},
  year={2022}
}

@incollection{ige2022enhancing,
  title={Enhancing border security and countering terrorism through computer vision: A field of artificial intelligence},
  author={Ige, Tosin and Kolade, Abosede and Kolade, Olukunle},
  booktitle={Proceedings of the Computational Methods in Systems and Software},
  pages={656--666},
  year={2022},
  publisher={Springer}
}

@article{ige2023adversarial,
  title={Adversarial sampling for fairness testing in deep neural network},
  author={Ige, Tosin and Marfo, William and Tonkinson, Justin and Adewale, Sikiru and Matti, Bolanle Hafiz},
  journal={arXiv preprint arXiv:2303.02874},
  year={2023}
}

@inproceedings{ige2022implementation,
  title={Implementation of data mining on a secure cloud computing over a web API using supervised machine learning algorithm},
  author={Ige, Tosin and Sikiru, Adewale},
  booktitle={Computer Science On-line Conference},
  pages={203--210},
  year={2022},
  organization={Springer}
}

@article{adewale2023encoder,
  title={Encoder-decoder based long short-term memory (lstm) model for video captioning},
  author={Adewale, Sikiru and Ige, Tosin and Matti, Bolanle Hafiz},
  journal={arXiv preprint arXiv:2401.02052},
  year={2023}
}

@article{okomayin2023ambient,
  title={Ambient technology \& intelligence},
  author={Okomayin, Amos and Ige, Tosin},
  journal={arXiv preprint arXiv:2305.10726},
  year={2023}
}

@article{Ige2024Exploiting,
  author    = {Ige, Tosin},
  title     = {Exploiting the In-Distribution Embedding Space with Deep Learning and Bayesian Inference for Detection and Classification of an Out-of-Distribution Malware (Extended Abstract)},
  journal   = {PhilArchive},
  year      = {2024},
  url       = {https://philarchive.org/rec/IGEETI},
  note      = {Extended abstract},
}

@article{Ige2025Impact,
  author    = {Ige, Tosin},
  title     = {Impact of Variation in Vector Space on the Performance of Machine and Deep Learning Models on an Out-of-Distribution Malware Attack Detection},
  journal   = {PhilArchive},
  year      = {2025},
  url       = {https://philarchive.org/rec/IGEIOV},
  note      = {Forthcoming IEEE Conference Proceeding},
}

\end{document}